\documentclass[aps,prl,preprint,superscriptaddress]{revtex4}

\usepackage{graphicx}

\begin{document}


\title{Uncollapsing the wave function}



\author{John Ashmead}
\email[]{akmed@voicenet.com}


\date{\today}

\newcommand{\sg}{Stern-Gerlach }
\newcommand{\seqn}{Schr\"odinger equation }

\begin{abstract}

The space quantization induced by a \sg experiment is normally explained 
	by invoking the ``collapse of the wave function.'' 
This is a rather mysterious idea;
	 it would be better to explain the \sg results without using it.

We re-analyze the \sg experiment using path integrals.  
We find if we model explicitly the finite width of the beam, 
	coherent interference within the beam itself provides the space quantization
	-- without need to invoke the collapse.  
If we insist on employing only wave functions 
	with the space and spin parts kept forcibly disentangled, 
	we recreate the need to invoke the collapse. 

The collapse-free approach makes more specific predictions about the shape and position 
	of the scattered beams; 
	if the interaction region has finite length, these may be testable.

Pending experimental disambiguation, 
	the chief arguments in favor of the collapse-free approach are that it is simpler 
	and less mysterious, 
	has no adjustable parameters, 
	and requires the invocation of no new forces.

\end{abstract}



\maketitle






\section{Introduction}

Quantum mechanics is over seventy-five years old 
	but remains something of a mystery. 
Much of the difficulty centers around the location of the Heisenberg cut, 
	the boundary between the quantum and classical realms. 
The main problem is the existence of such a boundary in the first place. 
Given that the world is emphatically quantum mechanical, 
	there should be no separate domain of competence for classical physics. 
All classical results should ultimately be explicated in quantum terms. 

The boundary between quantum and classical is normally defined as the point where the wave function ``collapses.''
The collapse takes place somewhere between the experiment and the observer 
	but is unusual among scientific concepts in that it may be found anywhere 
	but where it is being looked for. 
The doctrine of the collapse of the wave function may be traced back to Heisenberg\cite{Heisenberg:1930}, 
	with its mathematical formalization taking place in von Neumann's celebrated text\cite{vonNeumann:1955}. 
A review of the concept and its history is provided by Jammer\cite{Jammer:1966,Jammer:1974}.

While calculations using the collapse have been successful, 
	it creates a number of conceptual problems\cite{Ghose:1999}:
\begin{enumerate}
\item The ``measurement'' problem: 
	We should not have separate physics for measurement and process. 
	Measurements are themselves a process 
		and should be described by the same physics. 
	Whether we call something a ``measurement'' or a ``process'' 
		is a matter of labels; it should not affect the physics.
\item The ``event'' problem: 
	When does the collapse take place? 
	It must somewhere between the quantum and classical parts of the analysis, 
		but where in spacetime is the boundary located? 
	If we have $N$ observers each observing the other $N - 1$ 
		(and themselves for good measure) 
		are there $N^{2}$ collapses?
\item The ``preferred basis'' problem: 
	If we are looking at a collapse along the z axis, 
		then how did the system ``know'' it was to collapse along the z axis? 
	What if it became confused and collapsed along the y?
\end{enumerate}

In 1935 Einstein, Rosen, and Podolsky\cite{Einstein:1935} famously argued that quantum mechanics was incomplete 
in that it did not simultaneously give the values of complementary coordinates, say x and $p_x$.  
Bohr\cite{Bohr:1935}) famously refuted their objection by observing 
that a theory may only be \emph{required} to predict the results of experimentally measurable quantities 
(although it is free to introduce as many unmeasurable quantities as convenience suggests).
Only to the extent that  x and $p_x$ are simultaneously measurable 
	is it required that the theory predict them.

The preferred basis problem represents a different kind of incompleteness.
It implies a requirement that the quantum mechanic employ intuition 
	in deciding how to analyze a given experiment. 
This implies that quantum mechanics is not formally complete.
To put this in operational terms:
	there is not in general a way we could program software 
	to define the location of the Heisenberg cut. 
As Pearle\cite{Pearle:1994} puts it: the Copenhagen Interpretation 
	``is a good practical guide for working physicists. 
	Measurement by an apparatus is like great art, we know it when we see it
	-- and we have great artists who can bring it about. 
	But, a complete theoretical description? No.''

In the \sg experiment
	\cite{Gerlach:1922a,Gerlach:1922b,Gerlach:1922c}
	the collapse is assigned responsibility for the observed space quantization. 
The \sg experiment is a particularly good place to study quantization 
	because in this case many of the usual explanations of quantization are not available: 
\begin{enumerate}
\item Quantization is often supplied by a container of some kind: 
		a violin in the case of musical tones,
		a ``black box'' in the case of black body radiation, 
		a microwave cavity in the case of some mesoscopic cat experiments, 
		the central atomic potential in the case of Bohr-Sommerfeld quantization.  
	But in the case of the classic \sg experiment there is no container.
\item In other cases, quantized inputs naturally produce quantized outputs:
	 if an atomic system is going from one quantized energy level to another 
		the radiation it emits in so doing is automatically quantized as well.  
	But in the case of the \sg experiment the inputs are not quantized.
\item The program of decoherence
\cite{Namiki:1997,Omnes:1994a,Omnes:1997,Zeh:1996a,Zeh:2003} 
	has had significant success in explaining the tendency of macroscopic systems 
		to focus on a few pointer states. 
	It has been confirmed experimentally, e.g.
\cite{Raimond:1997,Brune:1996}.
	But decoherence is normally held to occur as a result of interaction with the environment
		\cite{Joos:1996,Joos:1999,Tegmark:2000b,Myatt:2000}. 
	In the case of the \sg experiment we may arrange, 
		in principle at least, 
		for the beam to have minimal interaction with the environment. 
\end{enumerate}
In the \sg effect, therefore, we see space quantization without obvious external explanation. 
To reproduce this quantization it would appear 
	we must rely solely on the collapse of the wave function, 
	naked and inescapable. 

One reasonable response to the doctrine of the collapse is to assume it is in some sense real 
	and to attempt to formalize it well enough to give it ``a local habitation and a name.'' 
This is the approach taken in the spontaneous localization (SL) 
	and continuous spontaneous localization (CSL) programs: 
	Ghirardi, Rimini, and Weber\cite{Ghirardi:1986}, Pearle\cite{Pearle:1989,Pearle:1994}, 
	and a good general review in Dickson\cite{Dickson:1998}. 
While these have not been experimentally confirmed
	\cite{Ghose:1999,Auletta:2000}, 
	their analyses have helped to highlight the conceptual difficulties associated with the collapse.

We ourselves were led to consider the problem from the point of view of Cramer's Transactional Interpretation
	\cite{Cramer:1986,Cramer:1988a}. 
We have found this interpretation fruitful but were troubled by the apparent need to invoke the collapse within it,
	at least to deal with the \sg experiment.
Given the natural connection between the Transactional Interpretation and path integrals 
	we tried using path integrals to analyze the problem.

We were a bit surprised to find that there was no need to invoke the collapse; 
	space quantization appeared spontaneously. 
Further examination made clear that this result 
	was not a side-effect of using the Transactional Interpretation or path integrals; 
	it resulted from using wave functions
	which modeled explicitly the width of the beam
	and whose spin and space components were entangled. 

When, as we will show, 
	we restrict the allowed wave functions 
	to include only those in which spin and space components are disentangled, 
	we recreate the necessity to invoke the collapse of the wave function. 
This is reasonable enough. 
As we know from the program of decoherence, 
	the boundary between the quantum and classical realms 
	may be defined by the point at which acceptable accuracy may be achieved 
	even if we restrict the solutions to those using only disentangled wave functions.

We will begin by examining [a very simple case of] 
	the \sg experiment in classical terms. 
Then we will employ path integrals to analyze the experiment. 
We will find that space quantization results 
	from coherent interference within the incident beam 
	-- without need for hand work. 
Then we will find if we insist the spin and space parts of the wave function be disentangled, 
	we reproduce the need for the collapse. 
With the location of the problem tentatively identified as ``premature disentanglement,'' 
	we will then examine possibilities for experimental test. 
These will depend largely on what one means by the ``collapse of the wave function,'' 
	a concept which resembles a politician in its ability to dodge sharp questions. 
The most obvious line of experimental attack is simply to note that the collapse-free approach 
	makes more specific predictions about the location and shape of the resulting wave function.

\section{Classical analysis of \sg experiment}

We begin with the classical problem. 
This provides us a well-defined starting point, 
	lets us establish notation, 
	and gives the value of the action along the classical path. 
A beam of uncharged spin 
	${{\left( {2m+1} \right)} \mathord{\left/ {\vphantom {{\left( {2m+1} \right)} 2}} \right. \kern-\nulldelimiterspace} 2}$
	particles is sent in the $y$ direction through an inhomogeneous magnetic field 
	${\vec B}$ 
	aligned along the z axis. 
We assume 
	${\vec B}$
	turns on and off abruptly. 
Since the \sg effect is produced only by the part of 
	${\vec B}$
	that varies with z we ignore the constant part of the field. 
To satisfy the requirement that 
	$\nabla \cdot \vec B=0$
	the magnetic field should include small x and y components, 
	but since we are only interested in points of principle we ignore these.

\begin{figure}
\includegraphics{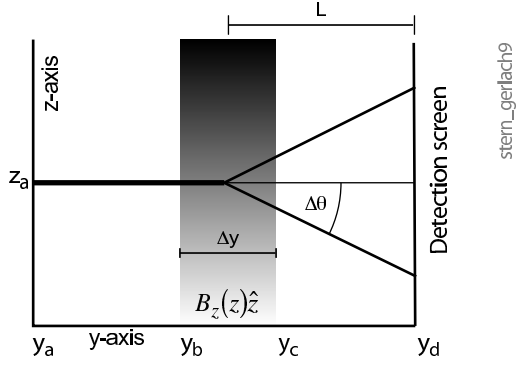}
\caption{\label{fig:seqn} {\sg experiment in schematic view}}
\end{figure}

If the particles have non-zero magnetic moment $\vec \mu$, their motion will be governed by the Hamiltonian
\begin{equation}
	H={{\vec p^2} \over {2m}}-\vec \mu \cdot \vec B
\end{equation}
	and they will experience a force
\begin{equation}
\vec F=\nabla \left( {\vec \mu \cdot \vec B} \right).
\end{equation}

For definiteness we will take the beam as going in the +y direction. 
It will appear at $y=y_a$, 
	enter the magnetic field at $y_b$,
	leave at $y_c$,
	and be detected at $y_d$.
We will denote the values of z at 
	$y_a$, $y_b$, $y_c$, and $y_d$
	as 
	$z_a (=0)$, $z_b$, $z_c$, and $z_d$
	with the corresponding times
	$t_a=t'$, $t_b$, $t_c$, and $t_d$.
We will assume that the beam is tightly focused 
	and that there are no significant $p_x$ or $p_z$ terms. 
We will take the magnetic field as going along the z axis 
	and having the form
\begin{eqnarray}
\vec B\left( {\vec x} \right)&\approx &B_z\hat z\hfill\cr
  B_z&\equiv &\left. {{{\partial B_z} \over {\partial z}}} \right|_{z=0}z\left( {\theta \left( {y-y_b} \right)-\theta \left( {y-y_c} \right)} \right)\hat z\hfill
\end{eqnarray}
where $\theta$ is the unit step function. 
The non-trivial equations of motion are
\begin{eqnarray}
\dot z &=&\left\{ {z,H} \right\}={{p_z} \over m}\hfill\cr
  \dot p_z &=&\left\{ {p_z,H} \right\}=\mu _z\left. {{{\partial B_z} \over {\partial z}}} \right|_{z=0}\hfill
\end{eqnarray}
in the interaction region, with
${\mu_z}$
the component of the spin along the z axis. 
The solutions are
\begin{eqnarray}
p_z\left( t \right)&=&p_z\left( {t_b} \right)+\left( {t-t_b} \right)\mu _z\left. {{{\partial B_z} \over {\partial z}}} \right|_{z=0}\hfill\cr
  z\left( t \right)&=&z_b+\left( {t-t_b} \right){{p_z\left( {t_b} \right)} \over m}
	+{{\mu _z} \over {2m}}\left. {{{\partial B_z} \over {\partial z}}} \right|_{z=0}\left( {t-t_b} \right)^2\hfill
\end{eqnarray}
Expressed in terms of the endpoints $t_b$, $z_b$, $t_c$, $z_c$
\begin{eqnarray}
p_z\left( t \right)&=&m{{z_c-z_b} \over {t_c-t_b}}+\mu _z\left. {{{\partial B_z} \over {\partial z}}} \right|_{z=0}\left( {t-{{t_c+t_b} \over 2}} \right)\hfill\cr
  \phantom{}&=&m{{\Delta z} \over {\Delta t}}+\mu _z\left. {{{\partial B_z} \over {\partial z}}} \right|_{z=0}\left( {t-\bar t} \right)\hfill\cr
  z\left( t \right)&=&z_b+{{\Delta z} \over {\Delta t}}\left( {t-t_b} \right)
	+{{\mu _z} \over {2m}}\left. {{{\partial B_z} \over {\partial z}}} \right|_{z=0}\left( {t-t_b} \right)\left( {t-t_c} \right)\hfill
\label{eq:classical:trajectory}
\end{eqnarray}

\begin{figure}
\includegraphics{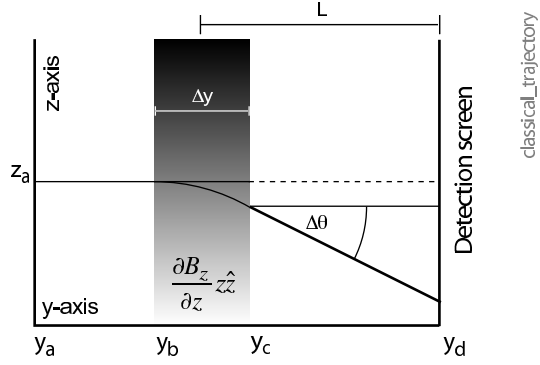}
\caption{\label{fig:classical:trajectory} {Classical trajectory in an \sg experiment}}
\end{figure}

To first order in the transit time through the interaction region, 
	the trajectory is a straight line through the interaction region 
	followed by a sharp kink up or down, 
	as given by the change in  $p_z$
\begin{equation}
\Delta p_z\equiv p_z\left( {t_c} \right)-p_z\left( {t_b} \right)=\mu _z\left. {{{\partial B_z} \over {\partial z}}} \right|_{z=0}\Delta t
\end{equation}
When $p_z(t_b)$  is zero the change in z is second order in the transit time
\begin{equation}
\Delta z={{p_z\left( {t_b} \right)} \over m}\Delta t+{{\Delta p_z} \over {2m}}\Delta t=\left. {{{\mu _z} \over {2m}}{{\partial B_z} \over {\partial z}}} \right|_{z=0}\left( {\Delta t} \right)^2
\end{equation}
The velocity is given by
\begin{equation}
v\equiv {{\sqrt {\left( {\vec x-\vec x_c} \right)^2+\left( {\vec x_c-\vec x_b} \right)^2+\left( {\vec x_b-\vec x'} \right)^2}} \over {t-t'}}
\end{equation}
Assuming that the angular deviation induced by the field is small, we have to lowest order
\begin{equation}
v\approx {{y-y'} \over {t-t'}}={{\Delta y} \over {\Delta t}}
\end{equation}	
\begin{equation}
t_b=v{{y_b-y'} \over {\Delta y}},\  t_c=v{{y_c-y'} \over {\Delta y}}
\end{equation}	
\begin{equation}
\Delta t={{\Delta y} \over v}=\Delta y{m \over {p_y}}
\end{equation}	
\begin{equation}
\Rightarrow \Delta p_z={{\Delta y} \over \nu }\mu _z\left. {{{\partial B_z} \over {\partial z}}} \right|_{z=0}
\end{equation}
We will usually be taking the limit as $\Delta y \rightarrow 0$ while the product
$\Delta y\left. {{{\partial B_z} \over {\partial z}}} \right|_{z=0}$ 
is constant. 
This will result in an angular change of direction of
\begin{equation}
\Delta \theta ={{\Delta p_z} \over {p_y}}={{m\Delta y} \over {p_y^2}}\left. {\mu _z{{\partial B_z} \over {\partial z}}} \right|_{z=0}
\end{equation}
This gives for $z_d$ the position at the detection screen
\begin{eqnarray}
z_d&\approx& L\Delta \theta &\hfill\cr
  L&\equiv&y_d-\left( {{{y_c+y_b} \over 2}} \right)=y_d-\bar y\hfill
\end{eqnarray}
In the case of the original \sg experiment 
	the magnetic moment of the silver atoms was given almost entirely by the magnetic moment of the unpaired electron
\begin{eqnarray}
\vec \mu &=&-{e \over {mc}}\vec s=-{{e\hbar } \over {2mc}}\vec \sigma =-\mu _b\vec \sigma &\hfill\cr
  \mu _b&\equiv& {{e\hbar } \over {2mc}}\hfill
\end{eqnarray}
With $\alpha$ and $\beta$ as the azimuthal and polar angles of the spin vector
\begin{eqnarray}
z_d&\approx& -L{{m\Delta y} \over {p_y^2}}\left. {\mu _b{{\partial B_z} \over {\partial z}}} \right|_{z=0}\left\langle {\vec \sigma } \right\rangle \cr
	&=&-L{{m\Delta y} \over {p_y^2}}\mu _b\left. {{{\partial B_z} \over {\partial z}}} \right|_{z=0}\cos \left( \beta  \right)\cr
	&=&-z_{\max }\cos \left( \beta  \right)\hfill\cr
  z_{\max }&\equiv& L{{m\Delta y} \over {p_y^2}}\mu _b\left. {{{\partial B_z} \over {\partial z}}} \right|_{z=0}\cr
	&=&v_zT\hfill\cr
  v_z&\equiv& {{\Delta y} \over {p_y}}\left. {\mu _b{{\partial B_z} \over {\partial z}}} \right|_{z=0}\hfill\cr
  T&=&L{m \over {p_y}}\hfill
\end{eqnarray}
	where $v_z$ is the post-interaction velocity in the z direction 
	and $T$ is the time it takes to travel from the interaction region to the detection plane. 
Isotropically distributed spins are distributed as
\begin{eqnarray}
{1 \over {4\pi }}\sin \left( \beta  \right)d\alpha d\beta &=&{1 \over 2}\sin \left( \beta  \right)d\beta \cr
	&=&{1 \over 2}d\left( {\cos \left( \beta  \right)} \right)\cr
	\Rightarrow p\left( {\cos \left( \beta  \right)} \right)&=&{1 \over 2}
\end{eqnarray}
giving a flat distribution
\begin{eqnarray}
p\left( {z_d} \right)&=&p\left( {\cos \left( \beta  \right)} \right){{d\cos \left( \beta  \right)} \over {dz_d}}\cr
	&=&{1 \over {2z_{\max }}}\left( {\theta \left( {z+z_{\max }} \right)-\theta \left( {z-z_{\max }} \right)} \right)
\end{eqnarray}

\begin{figure}
\includegraphics{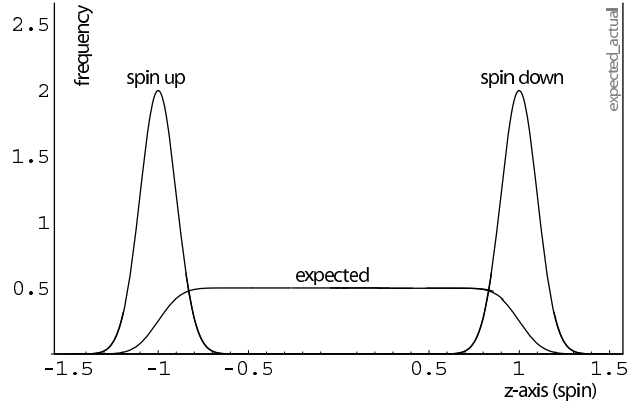}	
\caption{\label{fig:expected}Expected versus actual distributions}
\end{figure}

And of course what is seen is nothing like that: instead of a flat curve, 
	perhaps gently rounded at the ends (if the initial beam is really a Gaussian in z) 
	we see two relatively sharp spikes at the far ends of the classical distribution. 
It is as if the magnetic moment is given by $\mu_z=\mp \mu_b$
	with no intermediate values possible.

\section{Path integral analysis of \sg experiment}

We now calculate the results in the \sg experiment without invoking the collapse. 
We employ path integrals and limit ourselves to considering only a weak, well-localized magnetic field, as above. 
We will compute the propagator for a general spin-dependent potential, 
	specialize to the \sg case,
	then compute the wave function as a function of time 
	under the assumption that the initial wave function is a minimum uncertainty Gaussian. 
This will be enough to make the qualitative features clear. 
We will discuss possible improvements at the end of this section.

\subsection{Path integral calculation of propagator with spin}

We begin with the \seqn 
for a particle with spin $s$ represented by a spinor with   
	$M={{\left( {2s+1} \right)} \mathord{\left/ {\vphantom {{\left( {2s+1} \right)} 2}} \right. \kern-\nulldelimiterspace} 2}$
	components
	\footnote{We are following the treatment in Schulman\cite{Schulman:1981} 
		with the addition of a spin index and the elision of some mathematics.}, 
	with an $M$ by $M$ Hamiltonian $H$
\begin{equation}
i\hbar {\partial  \over {\partial t}}\left| \psi  \right\rangle =H\left| \psi  \right\rangle
\end{equation}
\begin{equation}
\Rightarrow \left| {\psi \left( t \right)} \right\rangle =\exp \left( {-{i \over \hbar }\int\limits_{t'}^t {dt''H\left( {t''} \right)}} \right)\left| {\psi \left( {t'} \right)} \right\rangle 
\end{equation}
\begin{equation}
\Rightarrow \left| {\psi \left( t \right)} \right\rangle =\left| {\vec x,s} \right\rangle \int {d\vec xd\vec x'\sum\limits_{\left\{ {ss'} \right\}} {\left\langle {\vec x,s} \right|\exp \left( {-{i \over \hbar }\int\limits_{t'}^t {dt''H\left( {t''} \right)}} \right)\left| {\vec x',s'} \right\rangle \left\langle {{\vec x',s'}} \mathrel{\left | {\vphantom {{\vec x',s'} {\psi \left( {t'} \right)}}} \right. \kern-\nulldelimiterspace} {{\psi \left( {t'} \right)}} \right\rangle }}
\label{eq:suitable:basis}
\end{equation}
where the $\left| {\vec x,s} \right\rangle $  represents some suitable basis in coordinate and spin space.
In the case we are interested in it may be broken into a kinetic energy part, $T$, diagonal in the spin component and a potential part, $V$, which in general mixes the spin components
\begin{equation}
H=T+V
\end{equation}
For time independent $H$ we may write the propagator $K$ as
\begin{eqnarray}
K\left( {t,\vec x,s;t',\vec x',s'} \right) 
	&\equiv& \left\langle {\vec x,s} \right|\exp \left( {-{i \over \hbar }\int\limits_{t'}^t {dt''H\left( {t''} \right)}} \right)\left| {\vec x',s'} \right\rangle \hfill\cr
  &=&\left\langle {\vec x,s} \right|\exp \left( {-i{{H\left( {t-t'} \right)} \mathord{\left/ {\vphantom {{H\left( {t-t'} \right)} \hbar }} \right. \kern-\nulldelimiterspace} \hbar }} \right)\left| {\vec x',s'} \right\rangle \hfill
\end{eqnarray}
We break the exponential up into a product of exponentials over infinitesimal time intervals
\begin{eqnarray}
K\left( {t,\vec x,s;t',\vec x',s'} \right) &=&\mathop {\lim }\limits_{N\to \infty }\left\langle {\vec x,s} \right|\left( {\exp \left( {-i{{\varepsilon T} \mathord{\left/ {\vphantom {{\varepsilon T} \hbar }} \right. \kern-\nulldelimiterspace} \hbar }} \right)\exp \left( {-i{{\varepsilon V} \mathord{\left/ {\vphantom {{\varepsilon V} \hbar }} \right. \kern-\nulldelimiterspace} \hbar }} \right)} \right)^N\left| {\vec x',s'} \right\rangle \hfill\cr
  \varepsilon &\equiv& {{t-t'} \over N}, \qquad t_j\equiv t'+j\varepsilon
\end{eqnarray}
The $t_0=t'$ and $t_N=t$ times are the starting and ending times respectively.
We insert a resolution of unity between each term in the product
\begin{eqnarray}
\lefteqn{K\left( {t,\vec x,s;t',\vec x',s'} \right)} \\
&=& {}\mathop {\lim }\limits_{N\to \infty }\int {d\vec x_1\ldots d\vec x_{N-1}\sum\limits_{\left\{ {s_1\ldots s_{N-1}} \right\}} {\prod\limits_{j=0}^{N-1} {\left\langle {\vec x_{j+1},s_{j+1}} \right|\exp \left( {-i{{\varepsilon T} \mathord{\left/ {\vphantom {{\varepsilon T} \hbar }} \right. \kern-\nulldelimiterspace} \hbar }} \right)\exp \left( {-i{{\varepsilon V} \mathord{\left/ {\vphantom {{\varepsilon V} \hbar }} \right. \kern-\nulldelimiterspace} \hbar }} \right)\left| {\vec x_j,s_j} \right\rangle }}} \nonumber
\end{eqnarray}
Sandwiched between two coordinate states the potential gives
\begin{eqnarray}
\left\langle {\vec x_{j+1},s_{j+1}} \right|V\left| {\vec x_j,s_j} \right\rangle
	&=&\delta \left( {\vec x_{j+1}-\vec x_j} \right)\left\langle {s_{j+1}} \right|V\left( {\vec x_j} \right)\left| {s_j} \right\rangle 
\end{eqnarray}
We deal with the kinetic energy part of the Green's function by inserting a momentum space resolution of unity
\begin{equation}
1=\int {d\vec p\sum\limits_{\left\{ s \right\}} {\left| {\vec p,s} \right\rangle \left\langle {\vec p,s} \right|}}
\end{equation}
on both sides of the kinetic energy term
\begin{eqnarray}
\lefteqn{\left\langle {\vec x_{j+1},s_{j+1}} \right|\exp \left( {-i{{\varepsilon T} \mathord{\left/ {\vphantom {{\varepsilon T} \hbar }} \right. \kern-\nulldelimiterspace} \hbar }} \right)\left| {\vec x_j,s_j} \right\rangle}
	 \\
  	& =&\int {d\vec p\sum\limits_{\left\{ s \right\}} {\int {d\vec p'\sum\limits_{\left\{ {s'} \right\}} {\left\langle {{\vec x_{j+1},s_{j+1}}} \mathrel{\left | {\vphantom {{\vec x_{j+1},s_{j+1}} {\vec p,s}}} \right. \kern-\nulldelimiterspace} {{\vec p,s}} \right\rangle \left\langle {\vec p,s} \right|\exp \left( {-i{{\varepsilon T} \mathord{\left/ {\vphantom {{\varepsilon T} \hbar }} \right. \kern-\nulldelimiterspace} \hbar }} \right)\left| {\vec p',s'} \right\rangle \left\langle {{\vec p',s'}} \mathrel{\left | {\vphantom {{\vec p',s'} {\vec x_j,s_j}}} \right. \kern-\nulldelimiterspace} {{\vec x_j,s_j}} \right\rangle }}}}\hfill\cr
  	&=&\int {d\vec p\sum\limits_{\left\{ s \right\}} {\int {d\vec p'\sum\limits_{\left\{ {s'} \right\}} {\left\langle {{\vec x_{j+1},s_{j+1}}} \mathrel{\left | {\vphantom {{\vec x_{j+1},s_{j+1}} {\vec p,s}}} \right. \kern-\nulldelimiterspace} {{\vec p,s}} \right\rangle \left\langle {\vec p,s} \right|\exp \left( {{{-i\varepsilon \vec p'^2} \over {2m\hbar }}} \right)\left| {\vec p',s'} \right\rangle \left\langle {{\vec p',s'}} \mathrel{\left | {\vphantom {{\vec p',s'} {\vec x_j,s_j}}} \right. \kern-\nulldelimiterspace} {{\vec x_j,s_j}} \right\rangle }}}}\nonumber
\end{eqnarray}
Since
\begin{equation}
\left\langle {{\vec p,s}} \mathrel{\left | {\vphantom {{\vec p,s} {\vec x,s'}}} \right. \kern-\nulldelimiterspace} {{\vec x,s'}} \right\rangle ={1 \over {\sqrt {\left( {2\pi \hbar } \right)^3}}}\exp \left( {{{-i\vec p\cdot \vec x} \over \hbar }} \right)\delta _{ss'}
\end{equation}
the matrix elements for the kinetic energy are
\begin{eqnarray}
\lefteqn{\left\langle {\vec x_{j+1},s_{j+1}} \right|\exp \left( {-i{{\varepsilon T} \mathord{\left/ {\vphantom {{\varepsilon T} \hbar }} \right. \kern-\nulldelimiterspace} \hbar }} \right)\left| {\vec x_j,s_j} \right\rangle } 
	\\
	& & ={{\delta _{s_{j+1}s_j}} \over {\left( {2\pi \hbar } \right)^3}}  
	\int {d\vec p}\exp \left( {{{{i\vec p\cdot \vec x_{j+1}} \over \hbar}
		-{{i\varepsilon \vec p^2} \over {2m\hbar }}} 
		-{{i\vec p\cdot \vec x_j} \over \hbar }}\right)
	\nonumber \\
	&& =\delta _{s_{j+1}s_j}\sqrt {{m \over {2\pi i\varepsilon \hbar }}}^3\exp \left( {{{i\varepsilon } \over \hbar }{m \over 2}\left( {{{\vec x_{j+1}-\vec x_j} \over \varepsilon }} \right)^2} \right)
	\nonumber
\end{eqnarray}
We can now write the path integral explicitly
\begin{eqnarray}
\lefteqn{K\left( {t,\vec x,s;t',\vec x',s'} \right)=}
	\nonumber \\
	& & \mathop {\lim }\limits_{N\to \infty }\int {d\vec x_1\ldots d\vec x_{N-1}\sum\limits_{\left\{ {s_1\ldots s_{N-1}} \right\}} {\sqrt {{m \over {2\pi i\varepsilon \hbar }}}^{3N}}}
	\nonumber \\
	& & \times \exp \left( {{{i\varepsilon } \over \hbar }\sum\limits_{j=0}^{N-1} {\left( {{m \over 2}\left( {{{\vec x_{j+1}-\vec x_j} \over \varepsilon }} \right)^2\mathord{\buildrel{\lower3pt\hbox{$\scriptscriptstyle\leftrightarrow$}}\over 1} -\mathord{\buildrel{\lower3pt\hbox{$\scriptscriptstyle\leftrightarrow$}}\over V} \left( {\vec x_j} \right)} \right)}} \right)
\end{eqnarray}

\subsection{Propagator for \sg experiment}

We specialize this to the case we are interested in: a spin 
${1 \mathord{\left/ {\vphantom {1 2}} \right. \kern-\nulldelimiterspace} 2}$
particle with a non-zero magnetic moment and no charge.
The \seqn is
\begin{eqnarray}
i\hbar {\partial  \over {\partial t}}\psi _i\left( {t,\vec x} \right)&=&-{{\hbar ^2} \over {2m}}\nabla ^2\psi _i\left( {t,\vec x} \right)-\sum\limits_{\left\{ j \right\}} {\vec \mu _{ij}\cdot \vec B\left( {\vec x} \right)\psi _j\left( {t,\vec x} \right)}
	\nonumber \\
  \left\{ j \right\}&=&\left\{ {-{1 \mathord{\left/ {\vphantom {1 {2,{1 \mathord{\left/ {\vphantom {1 2}} \right. \kern-\nulldelimiterspace} 2}}}} \right. \kern-\nulldelimiterspace} {2,{1 \mathord{\left/ {\vphantom {1 2}} \right. \kern-\nulldelimiterspace} 2}}}} \right\}
\end{eqnarray}

Since the magnetic field is pointing along the z axis, 
	we select the representation in which the spin is diagonal along the z axis 
	and write this as
\begin{eqnarray}
i\hbar {\partial  \over {\partial t}}\psi _i\left( {t,\vec x} \right)&=&-{{\hbar ^2} \over {2m}}\vec \nabla ^2\psi _i\left( {t,\vec x} \right) + \sum\limits_{\left\{ j \right\}} {\mu _bB_z\left( {\vec x} \right)\left( {\sigma _z} \right)_{ij}\psi _j\left( {t,\vec x} \right)}
\label{eq:spin}
\end{eqnarray}
We are \emph{not} insisting that the wave function be \emph{either} spin up or spin down -- that would be to invoke the collapse.
We continue to see the wave function as including \emph{both} spin up and spin down parts, but treat each part separately for mathematical convenience.
The up and down components of the wave function will propagate like separate wave functions.
At the end of the calculation we will join the two halves together.
We have
\begin{equation}
\left\langle {\vec x_{j+1},s_{j+1}} \right|\exp \left( {-{{i\varepsilon V} \over \hbar }} \right)\left| {\vec x_j,s_j} \right\rangle =\delta \left( {\vec x_{j+1}-\vec x_j} \right)\left( {\matrix{{\exp \left( {-{{i\varepsilon \mu _bB_z\left( {\vec x_j} \right)} \over \hbar }} \right)}&0\cr
0&{\exp \left( {{{i\varepsilon \mu _bB_z\left( {\vec x_j} \right)} \over \hbar }} \right)}\cr
}} \right)
\end{equation}
Since the off-diagonal elements are zero, the sums over the $s_j$ telescope; 
we may therefore write the propagator separately for the spin up and down parts
\begin{eqnarray}
K_\pm \left( {t,\vec x;t',\vec x'} \right)=&&
	 \\
 &  \mathop {\lim }\limits_{N\to \infty }\int {d\vec x_1\ldots d\vec x_{N-1}\sqrt {{m \over {2\pi i\varepsilon \hbar }}}^{3N}\exp \left( {{{i\varepsilon } \over \hbar }\sum\limits_{j=0}^{N-1} {\left( {{m \over 2}\left( {{{\vec x_{j+1}-\vec x_j} \over \varepsilon }} \right)^2\mp \mu _bB_z\left( {\vec x_j} \right)} \right)}} \right)} &
	\nonumber
\end{eqnarray}
So we may therefore write $\psi$ as
\begin{equation}
\psi \left( {t,\vec x} \right)=\left( {\matrix{{\int {d\vec x'K_+\left( {t,\vec x;t',\vec x'} \right)\psi _+\left( {t',\vec x'} \right)}}\cr
{\int {d\vec x'K_-\left( {t,\vec x;t',\vec x'} \right)\psi _-\left( {t',\vec x'} \right)}}\cr
}} \right)
\end{equation}
Essentially we have taken the coupled equations in 
$\psi _{\pm {1 \mathord{\left/ {\vphantom {1 2}} \right. \kern-\nulldelimiterspace} 2}}$
and reduced them to two independent equations.

\begin{figure}
\includegraphics{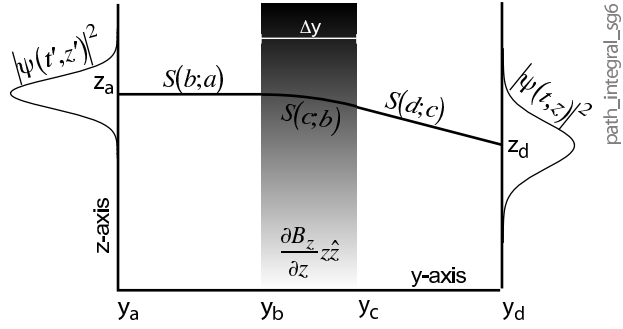}	
\caption{\label{fig: path:integral:sg} {Path integral calculation for \sg}}
\end{figure}

Per Schulman\cite{Schulman:1981}, we approximate the propagator as
\begin{equation}
K\left( {t,t'} \right)\approx \sqrt {{m \over {2\pi i\hbar f_{cl}\left( {t,t'} \right)}}}\exp \left( {{i \over \hbar }S_{cl}\left( {t,t'} \right)} \right)
\end{equation}
	where $S_{cl}(t,t')$  is the classical action from  $t'$ to $t$ and $f_{cl}(t,t')$  satisfies the equations
	(specialized from Schulman's treatment to our case)
\begin{eqnarray}
m{{\partial ^2f_{cl}\left( {t,t'} \right)} \over {\partial t^2}}=0,\qquad
  f_{cl}\left( {t',t'} \right)=0,\qquad
  \left. {{{\partial f_{cl}\left( {t,t'} \right)} \over {\partial t}}} \right|_{t=t'}=1\hfill
\end{eqnarray}
which implies
\begin{equation}
f_{cl}\left( {t,t'} \right)=t-t'
\end{equation}
If the Lagrangian is no worse than quadratic in the integration variables, this is exact.
For the action we have
\begin{eqnarray}
S\left( {t,\vec x;t',\vec x'} \right)&=&
	\int\limits_{t'}^{t} {{{p_x^2\left( t \right)} \over {2m}}+{{p_y^2\left( t \right)} \over {2m}}dt}
	+\int\limits_{t_c}^{t} {{{p_z^2\left( t \right)} \over {2m}}dt}\cr
	&+&\int\limits_{t_b}^{t_c} {\left( {{{p_z^2\left( t \right)} \over {2m}}+\mu _z{{\partial B_z} \over {\partial z}}z\left( t \right)} \right)dt}\cr
	&+&\int\limits_{t'}^{t_b} {{{p_z^2\left( t \right)} \over {2m}}dt}
\end{eqnarray}
Substituting in the classical trajectory Eq. 
(\ref{eq:classical:trajectory}) 
we get
\begin{eqnarray}
S\left( {t,\vec x;t',\vec x'} \right)&=&{m \over 2}{{\left( {x-x'} \right)^2} \over {t-t'}}
	+{m \over 2}{{\left( {y-y'} \right)^2} \over {t-t'}}
	+{m \over 2}{{\left( {z-z_c} \right)^2} \over {t-t_c}}\cr
	&+&{{m\left( {z_c-z_b} \right)^2} \over {2\left( {t_c-t_b} \right)}}
  	+\mu _z{{\partial B_z} \over {\partial z}}{{z_c+z_b} \over 2}\left( {t_c-t_b} \right) 
	-{1 \over {24m}}\left( {\mu _z{{\partial B_z} \over {\partial z}}} \right)^2\left( {t_c-t_b} \right)^3\cr
	&+&{m \over 2}{{\left( {z_b-z'} \right)^2} \over {t_b-t'}}\hfill
\end{eqnarray}


We are primarily interested in  the value of the z part of the action.
Dropping terms second order in the magnetic field we have
\begin{eqnarray}
S^{\left( z \right)}\left( {t,\vec x;t',\vec x'} \right)&=&{m \over 2}{{\left( {z-z_c} \right)^2} \over {t-t_c}}
	+{m \over 2}{{\left( {z_c-z_b} \right)^2} \over {t_c-t_b}}
	+\mu _z\left. {{{\partial B_z} \over {\partial z}}} \right|_{z=0}{{\Delta y} \over \nu }{{z_c+z_b} \over 2}
	+{m \over 2}{{\left( {z_b-z'} \right)^2} \over {t_b-t'}}\cr
  &=&{m \over 2}{{\left( {z-z_c} \right)^2} \over {t-t_c}}
	+{m \over 2}{{\left( {z_c-z_b} \right)^2} \over {t_c-t_b}}
	\mp mv_z{{z_c+z_b} \over 2}+{m \over 2}{{\left( {z_b-z'} \right)^2} \over {t_b-t'}}.
\end{eqnarray}
Then
\begin{eqnarray}
K^{\left( z \right)}\left( {t,z;t',z'} \right)&=&\sqrt {{m \over {2\pi i\hbar \left( {t-t_c} \right)}}}\sqrt {{m \over {2\pi i\hbar \left( {t_c-t_b} \right)}}}\sqrt {{m \over {2\pi i\hbar \left( {t_b-t'} \right)}}}\int {dz_c}dz_b\hfill\\
  	&&\times
		\exp \left( {{{im} \over {2\hbar }}{{\left( {z-z_c} \right)^2} \over {t-t_c}}} \right)
		\exp \left( {{{im} \over {2\hbar }}{{\left( {z_c-z_b} \right)^2} \over {t_c-t_b}}\mp i{{mv_z} \over \hbar }{{z_b+z_c} \over 2}} \right)\nonumber\\
	&&\times	\exp \left( {{{im} \over {2\hbar }}{{\left( {z_b-z'} \right)^2} \over {t_b-t'}}} \right) \nonumber
\end{eqnarray}
Recalling $t_c=\bar t+{{\Delta t} \mathord{\left/ {\vphantom {{\Delta t} 2}} \right. \kern-\nulldelimiterspace} 2},\  t_b=\bar t-{{\Delta t} \mathord{\left/ {\vphantom {{\Delta t} 2}} \right. \kern-\nulldelimiterspace} 2}$
and discarding terms of order $\Delta t$, we get
\begin{eqnarray}
K^{\left( z \right)}\left( {t,z;t',z'} \right)&=&\sqrt {{m \over {2\pi i\hbar \left( {t-t'} \right)}}}
	\exp \left( {{im} \over {2\hbar }}{{\left( {z-z'} \right)^2} \over {t-t'}} \right)\\
  & &\times  \exp \left( {\mp {{imv_z} \over \hbar }{{z\left( {\bar t-t'} \right)+z'\left( {t-\bar t} \right)} \over {t-t'}}-{{imv_z^2} \over {2\hbar }}{{\left( {t-\bar t} \right)\left( {\bar t-t'} \right)} \over {t-t'}}} \right) \nonumber
\end{eqnarray}
Again dropping terms second order in the magnetic field
\begin{equation}
K\left( {t,\vec x;t',\vec x'} \right)=K^{\left( {free} \right)}\left( {t,\vec x;t',\vec x'} \right)\exp \left( {\mp {{imv_z} \over \hbar }{{z\left( {\bar t-t'} \right)+z'\left( {t-\bar t} \right)} \over {t-t'}}} \right)
\label{eq:magnetic:kernel}
\end{equation}

In other words the kernel looks as if part of the change in $p_z$ takes effect at the beginning of the interaction region and the rest at the end (provided the interaction region is not too long).
If the interaction region is positioned 20\% of the way from $t'$ to $t$, then 80\% of the change will appear to have taken place at $t_b$ , 20\% at  $t_c$, which makes sense.

We have taken considerable advantage of the assumed simplicity 
	--- weak, constant in time, sharply delimited in space 
	--- of the magnetic potential.
If we were facing a more realistic spin-dependent potential, 
	say a time-dependent one, 
	it would make sense to compute the path integral using a variational approach.
At each time slice we would have a 2 by 2 or $M$ by $M$ matrix to diagonalize.
The matrix rotations required to do the diagonalizations 
	would define a time-dependent ``classical'' solution between the specified endpoints.
Approximating the integrals by Gaussians 
	would give the first quantum corrections.
Essentially we would expect to get the Coherent Internal States solutions 
	of Cruz-Barrios and G—mez-Camacho
	\cite{Cruz-Barrios:1998,Cruz-Barrios:2001}, 
	if from a slightly different direction.

\subsection{Wave function in \sg experiment}

\newcommand{\phinorm}{
{1 \over {\left( {\pi \sigma ^2} \right)^{3/4}}}
\sqrt {{1 \over {f\left( {t-t'} \right)}}}^3
}

\newcommand{\phiwave}{
\exp \left( {-{{\left( {\vec x-\vec x_a} \right)^2} \over {2\sigma ^2f\left( {t-t'} \right)}}+i{{k_y\left( {y-y_a} \right)} \over {f\left( {t-t'} \right)}}} \right)
}

\newcommand{\pnorm}{
\sqrt {{1 \over {\pi \sigma ^2\left| {f\left( {t-t'} \right)} \right|^2}}}^3
}

We take the input wave function as a minimum uncertainty wave packet, a Gaussian with initial direction along the y axis at  $t'$
\begin{eqnarray}
\varphi \left( {t',\vec x'} \right)&\equiv&{1 \over {\left( {\pi \sigma ^2} \right)^{3/4}}}\exp \left( {-{{\left( {\vec x'-\vec x_a} \right)} \over {2\sigma ^2}}^2+i\vec k\cdot \left( {\vec x'-\vec x_a} \right)} \right),\  \vec x_a\equiv \left( {0,y_a,0} \right),\  \vec k\equiv \left( {0,\hbar k_y,0} \right)\hfill\cr
  \psi \left( {t',\vec x'} \right)&\equiv& \left( {\matrix{{\varphi \left( {t',\vec x'} \right)\chi _+}\cr
{\varphi \left( {t',\vec x'} \right)\chi _-}\cr
}} \right)\hfill
\end{eqnarray}
Applying the kernel (\ref{eq:magnetic:kernel}) to this and integrating over  $\vec x'$
\begin{eqnarray}
\varphi _\pm \left( {t,\vec x} \right)&=&\phinorm  \phiwave\cr
	&&\times \exp \left( {\mp i{{mv_zf\left( {\bar t-t'} \right)z} \over {\hbar f\left( {t-t'} \right)}}-{{i\hbar \left( {k_y^2+\left( {{{mv_z} \over \hbar }{{t-\bar t} \over {t-t'}}} \right)^2} \right)\left( {t-t'} \right)} \over {2mf\left( {t-t'} \right)}}} \right)
\cr
  	f\left( t \right)&\equiv& 1+{{i\hbar t} \over {m\sigma ^2}}\hfill
\label{eq:wave}
\end{eqnarray}
This gives a probability of finding a particle of either spin as
\begin{eqnarray}
p\left( {t,\vec x} \right)&=&\pnorm \exp \left( {-{{x^2+\left( {y-\left( {y_a+v_y\left( {t-t'} \right)} \right)} \right)^2} \over {\sigma ^2\left| {f\left( {t-t'} \right)} \right|^2}}} \right)\hfill\cr
&&\times \left( {\exp \left( {-{{\left( {z+v_z\left( {t-\bar t} \right)} \right)^2} \over {\sigma ^2\left| {f\left( {t-t'} \right)} \right|^2}}} \right)\left| {\chi _+} \right|^2+\exp \left( {-{{\left( {z-v_z\left( {t-\bar t} \right)} \right)^2} \over {\sigma ^2\left| {f\left( {t-t'} \right)} \right|^2}}} \right)\left| {\chi _-} \right|^2} \right)\cr
\left| {f\left( t \right)} \right|^2&=&1+\left| {{{\hbar t} \over {m\sigma ^2}}} \right|^2
\end{eqnarray}
The ${\left|f(t)\right|}^2$ terms tells us 
	the smaller the initial width $\sigma$ of the beam 
	the faster it diverges, 
	as we would expect from the uncertainty principle.

In this last equation we see the expected two-humped distribution, with peaks at 
$z=\mp v_z\left( {t-\bar t} \right)$. 
The difference between this and the collapsed version is subtle but definite: 
in the collapsed version the wave function is \emph{either} spin up or spin down: 
there is no relative phase information between the two.
Here the wave function is \emph{both} spin up and spin down.
The relative phase information is preserved.
There is an exponentially small spin-up component to be found in the spin-down direction and vice versa.
The collapse-free density matrix (looking at the z-dependence only) is given by
\begin{equation}
\rho ^{\left( {collapse-free} \right)}\left( z \right)=\left( {\matrix{{\left| {\varphi _+\left( z \right)} \right|^2\left| {\chi _+} \right|^2}&{\varphi _+^*\left( z \right)\varphi _-\left( z \right)\chi _+^*\chi _-}\cr
{\varphi _-^*\left( z \right)\varphi _+\left( z \right)\chi _-^*\chi _+}&{\left| {\varphi _-\left( z \right)} \right|^2\left| {\chi _-} \right|^2}\cr
}} \right)
\end{equation}
The collapsed wave function has the density matrix
\begin{equation}
\rho ^{\left( {collapsed} \right)}\left( z \right)=\left( {\matrix{{\left| {\varphi _+\left( z \right)} \right|^2\left| {\chi _+} \right|^2}&0\cr
0&{\left| {\varphi _-\left( z \right)} \right|^2\left| {\chi _-} \right|^2}\cr
}} \right)
\end{equation}
The collapse-free and collapsed versions have the same trace and therefore the same probability for detection at the detection plane.

The off-diagonal terms are different.
But as $\phi_+$ is centered on the spin-up beam and $\phi_-$ is centered on the spin-down beam, the amount of spatial overlap between the two is exponentially small.
Except near the beginning of the interaction region, the off-diagonal terms will be nearly zero.
This implies that as a practical matter it would be tricky to distinguish between the two.
We therefore understand why, even if the collapse-free approach is in fact correct, the invocation of the collapse still gives good results; the two density matrices are almost identical except near the interaction region.
Invoking the collapse of the wave function is unnecessary, but almost harmless.
The major difference between the two approaches is (obviously) that the collapse-free approach does not invoke the collapse.
Since space quantization still appears, it would seem that the step ``collapse the wave function'' is not necessary to produce that.

Still, if we have lost a collapse, 
we have gained a catastrophe.
If we monitor  
$p\left( {t,\vec x} \right)\equiv \left| {\psi \left( {t,\vec x} \right)} \right|^2$
as it moves through the interaction region it will acquire at the beginning a small dimple in the middle, as the two spins begin to separate.
If we take the centroids of  
$\left| {\varphi _+} \right|^2$
and  
$\left| {\varphi _-} \right|^2$
as defining the classical paths, on encountering the magnetic field the initially single path splits into two.
In the language of catastrophe theory\cite{Arnold:1986} this is a cusp or bifurcation catastrophe.

There is still a sense in which the collapse still takes place.
When the particle is ultimately detected it will be found definitely at one location, within one of the two Gaussian beams.
This takes place as far past the interaction region as we have patience to wait for the particle.
It creates no ambiguity as to what basis to use.

\section{Disentangling the collapse}

If we are correct in thinking that the standard formalism, vigorously applied, 
	will generate space quantization without invocation of the collapse, 
	then we would like to understand what changes in that formalism 
	might recreate the necessity for the collapse, the better to locate the crux.
Our suspicion is that the key problem is the inadvertent projection of classical ideas onto the quantum realm.

It is conventional in discussions of \sg experiments to talk about the spatial and spin parts as if they were distinct.
The spatial part is treated as a kind of carrier wave; the spin as the signal impressed on it.
The schematic diagram of the \sg experiment will almost inevitably look like a road map, with forks at each piece of experimental apparatus.
It is as if the particles are very small cars, each with a propeller strapped on top.
Mostly the cars travel along fixed roadways, the pathways determined by the collapse of the wave function.
At critical junctions, their drivers check which way the propeller is pointing at the instant and are so guided, electing the spin up or spin down or the spin left or spin right pathway accordingly.
While the actual formalism treats spin and spatial components as entangled, the more informal discussions that surround each set of formal steps often treat spin and space as if they were separate.

The \seqn has a (free) spatial part and a spin-mixing part.
We will therefore attempt to solve this using a deliberately incorrect ansatz 
	(modeled on the lines of the intuitive picture). 
We will write the wave functions as composed of disentangled space and spin parts
\begin{equation}
\psi \left( {t,\vec x} \right)=\varphi \left( {t,\vec x} \right)\chi \left( t \right)=\left( {\matrix{{\varphi \left( {t,\vec x} \right)\chi _+\left( t \right)}\cr
{\varphi \left( {t,\vec x} \right)\chi _-\left( t \right)}\cr
}} \right)
\end{equation}
In the case of the path integral calculation, the initial wave function had, by assumption, this form.
But an effect of passing through the magnetic field was to turn it into a more complex wave function with different spatial dependence in the spin up and spin down parts.

With this ansatz, Eq. (\ref{eq:spin}) gives
\begin{eqnarray}
\hbar {\partial  \over {\partial t}}\varphi \left( {t,\vec x} \right)\chi _i\left( t \right)=	-{{\hbar ^2} \over {2m}}\nabla ^2\varphi \left( {t,\vec x} \right)\chi _i\left( t \right)-\vec \mu _{ij}\cdot \vec B\left( {\vec x} \right)\varphi \left( {t,\vec x} \right)\chi _j\left( t \right) 
\end{eqnarray}
If the interaction term did not mix the spin and space variables 
	the ansatz would work. 
If the separation of variables were possible, 
	the averages would be the separation constants.
Using the averages is therefore a reasonable way to handle our incorrect ansatz.  
If we take the average of the spin part 
	as giving the effect of $\mu$ on $\phi$ 
	and the average of the magnetic field 
	as giving the effect of $\phi$ on $\chi$ we get
\begin{eqnarray}
i\hbar \left( {{\partial  \over {\partial t}}\varphi \left( {t,\vec x} \right)} \right)
	&=&-{{\hbar ^2} \over {2m}}\left( {\nabla ^2\varphi \left( {t,\vec x} \right)} \right)-\left\langle {\vec \mu \left( t \right)} \right\rangle \cdot \vec B\left( {\vec x} \right)\varphi \left( {t,\vec x} \right)\hfill\cr
  i\hbar {\partial  \over {\partial t}}\chi _i\left( t \right)&=&-\vec \mu _{ij}\cdot \left\langle {\vec B\left( {t} \right)} \right\rangle \chi _j\left( t \right)
\end{eqnarray}
where
\begin{eqnarray}
\left\langle {\vec \mu \left( t \right)} \right\rangle &\equiv& \left\langle \chi  \right|\vec \mu \left| \chi  \right\rangle =-\mu _b\sum\limits_{\left\{ {i,j} \right\}} {\chi _i^\dagger \left( t \right)\vec \sigma _{ij}\chi _j\left( t \right)} \\
  \left\langle {\vec B\left( t \right)} \right\rangle &\equiv& \left\langle \varphi  \right|\vec B\left| \varphi  \right\rangle =\int {d\vec x\varphi ^\dagger \left( {t,\vec x} \right)\vec B\left( {\vec x} \right)\varphi \left( {t,\vec x} \right)} \nonumber
\end{eqnarray}
The initial value for  
$\left\langle {\vec \mu } \right\rangle $
 is determined by the initial value for the spin, which may point in any direction.
To lowest order in $B$, $\chi$ is a constant, therefore 
$\left\langle {\vec \mu \left( t \right)} \right\rangle =\left\langle {\vec \mu \left( {t'} \right)} \right\rangle $
is, giving
\begin{eqnarray}
i\hbar {\partial  \over {\partial t}}\varphi \left( {t,\vec x} \right)	=-{{\hbar ^2} \over {2m}}\nabla ^2\varphi \left( {t,\vec x} \right)-\left\langle {\vec \mu \left( {t'} \right)} \right\rangle \cdot \vec B\left( {\vec x} \right)\varphi \left( {t,\vec x} \right)
\end{eqnarray}

At this point the equation for  $\phi$ is what we had in the last section for each component of $\psi$, except that the 
$\vec \mu \cdot \vec B$
will span a range of values. 
Therefore for fixed $\vec \mu$  
	we get the wave function in Eq. (\ref{eq:wave}) with the replacement 
$\mp v_z\to \left\langle {\mu _z} \right\rangle {{v_z} \over {\mu _b}}$
\begin{eqnarray}
\varphi \left( {t,\vec x} \right)&=&
	\phinorm \phiwave
	\nonumber \\
	&&\times\exp \left( {i{{\left\langle {\mu _z} \right\rangle } \over {\mu _b}}{{mv_zf\left( {\bar t-t'} \right)z} \over {f\left( {t-t'} \right)}}-{{i\hbar k_y^2t} \over {2mf\left( {t-t'} \right)}}} \right)
\end{eqnarray}
with probability
\begin{equation}
p\left( {t,\vec x} \right)=\pnorm \exp \left( {-{{x^2+\left( {y-\left( {y_a+v_y\left( {t-t'} \right)} \right)} \right)^2+\left( {z-{{\left\langle {\mu _z} \right\rangle } \over {\mu _b}}v_z\left( {t-\bar t} \right)} \right)^2} \over {\sigma ^2\left| {f\left( t \right)} \right|^2}}} \right)
\end{equation}
This is a Gaussian centered at 
$z={{\left\langle {\mu _z} \right\rangle } \over {\mu _b}}v_z\left( {t-\bar t} \right)$, 
which spans the range from the spin up to the spin down values.
For isotropic spin distributions it will be distributed as  $cos(\beta)$.
To get the full probability distribution we convolute the Gaussian with the flat initial distribution for the probability, 
giving the ``expected'' classical distribution of Fig.~\ref{fig:expected}.

In other words, if we assume the spin and space components are disentangled we reproduce the classical prediction.
This makes sense.
The program of decoherence suggests 
	the boundary between classical and quantum modes of analysis 
	may be defined by the point 
	at which it becomes acceptable 
	to replace entangled quantum wave functions 
	with disentangled classical billiard balls and their cuesticks
	\footnote{In the program of decoherence a second boundary, 
		between reversible and irreversible processes, 
		is crossed at the same point}.

\section{Experimental predictions}

\begin{figure}
\includegraphics{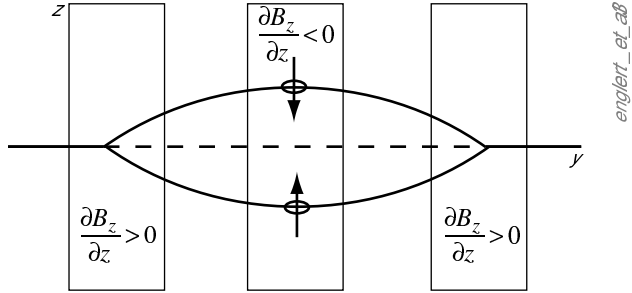}	
\caption{\label{fig: englert:et:al} {Schematic \sg with recombination}}
\end{figure}

The chief advantage of the collapse-free approach 
	is that it is simpler and less mysterious.
It appears that if we model the finite width of the beam explicitly 
	and refrain from invoking the collapse, but otherwise proceed as usual, 
	we will see space quantization appear spontaneously
	as a result of coherent self-interference effects within the beam.
We would like to identify experimental differences between the two approaches as well.

There are two detectible differences between the collapsed and the collapse-free approaches.
One is that in the collapse-free approach 
	we take the off-diagonal elements of the density matrix as non-zero.
This implies that the two (or more) separated beams are still in principle coherent 
	and if brought together again, would be able to interfere with one another.
For instance, if we start with a spin +x beam, split it in the z direction, 
	recombine the two beams, and check the spin along the x direction, 
	we should find it still in a +x state.
(See Fig. \ref{fig: englert:et:al} for a very schematic illustration.)
This line of attack as been explored by Englert, Scully, and Schwinger
	 in their series of papers
	\cite{Englert:1988,Schwinger:1988,Scully:1989a}:  
	``Is Spin Coherence Like Humpty Dumpty?'' 
They show that this is possible, 
	although not particularly easy.
Maintaining phase coherence while separating and recombining the beams is non-trivial.

\begin{figure}
\includegraphics{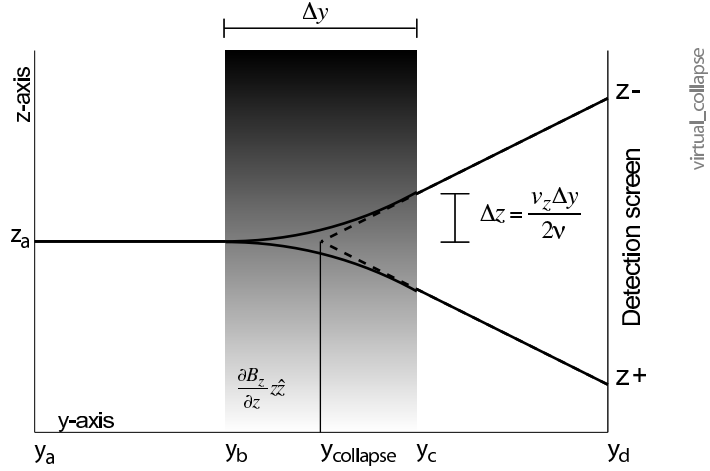}	
\caption{\label{fig:virtual:collapse} {Location of collapse}}
\end{figure}

The other (in principle) detectible difference is that the location and shape of the final wave function 
	is slightly different in the collapsed and collapse-free cases. 
The most obvious difference comes from the finite length of the interaction region. 
In the collapsed case the beam encounters the magnetic field 
	and at some point collapses and heads either up or down, 
	as its spin determines. 
If the interaction region is $\Delta y$ long, with a field exerting force 
$m\mu_z{{\partial B_z} \over {\partial z}}$
on a particle traveling with speed $\nu$ (see Fig. \ref{fig:virtual:collapse}), 
we will see a change of direction given by 
$v_z$.
Per Eq. (\ref{eq:classical:trajectory}) we have for the displacement in the z direction
\begin{equation}
\Delta z\equiv z_c-z_a=z_c={{v_z} \over 2}\Delta t={{v_z} \over 2}{{\Delta y} \over \nu }
\end{equation}

And we have that the centroid of the wave function goes in the direction  $\vec v=(0,v,v_z)$ after this. 
This lets us backtrack the resultant course, giving an effective location for the point of collapse as
\begin{equation}
y_{collapse}=y_c-{v \over {v_z}}\Delta z=y_c-{{\Delta y} \over 2}={{y_b+y_c} \over 2}=\bar y
\end{equation}
so we have a well-defined location for the collapse, as the center of the interaction region. 
There is no fundamental requirement that $y_{collapse}$
	be exactly the average of the start and end of the interaction region. 
If we had been less cavalier in our approximations or looked at a more complex situation 
	we would see corresponding adjustments in the value of $y_{collapse}$.
The collapse-free approach further predicts that $z_d$, 
	the centroid of the detected beam, 
	will be at  
$z_d=\mp {{v_z} \over v}\left( {y_d-y_{collapse}} \right)$. 

Ironically therefore, it is the collapse-free approach 
	which gives a well-defined location for the collapse of the wave function
\footnote{It would therefore more accurate to say not that the collapse-free 
	and the collapsed approaches give different predictions for $y_{collapse}$
	but rather that the collapse free approach gives a definite prediction
	and the collapsed approach merely implies that it must be somewhere 
	in the interaction region, 
	in our nomenclature somewhere between $y_b$ and $y_c$.}.
It is at the intersection of the backtracked output paths, 
	with the paths defined by their centroids. 
The thus defined collapse is not in general actually on either of the paths itself. 
It is a virtual location (appropriately enough) not a real one. 
In general, in any particular experimental case, 
	the collapse-free approach will select one specific location as the point of collapse
	\footnote{Unless we use such diabolically twisted magnetic fields 
		that the backtracks from the output beams have no point of intersection.}.
And this location will in turn define the expected target points. If the centroids are higher or lower than this, the collapse-free hypothesis is falsified.

The view we are led to by this is that 
	the collapse of the wave function represents a kind of approximation scheme. 
It gives a stick figure representation of a more detailed calculation. 
It cannot itself define the location of the collapse, but it does give a qualitatively correct representation. 
To get a correct analysis the path integrals used above or some other technique must be employed. 
So long as the width of the beam is modeled explicitly and entangled wave functions are allowed, space quantization should be seen.

We have used the simplest possible treatment consistent with reproducing the qualitative features. 
A number of possible improvements in the treatment of the input wave, magnetic field, and detectors are obvious. 
We could model the input wave not as an elementary Gaussian but an arbitrary wave form, perhaps modeled as a sum over Gaussian wavelets as described in Kaiser\cite{Kaiser:1994}. 
We could go to considerably greater lengths in describing the magnetic field: use second derivative and higher terms in z (or for that matter, include the constant part), include the necessarily present x and y components, turn the field on and off a bit less abruptly, allow the field to vary in time, model the magnetic field itself in quantum mechanical terms. 
And we could model the interaction with the detector wave functions explicitly, perhaps treating them as [rather narrow] Gaussians.

\section{Discussion}

We conclude that we do not need to invoke the collapse of the wave function 
	to explain the space quantization seen in the \sg experiment. 
Provided that we model the apparatus in sufficient detail, 
	space quantization will appear as a result of coherent interference within the beam.

The term ``in sufficient detail'' means that we explicitly include the finite width of the beam, 
refrain from limiting the set of allowed wave functions to only those with space and spin components disentangled, 
and model the inhomogeneity in the magnetic field to at least first order in z. 
The minimum width of the beam is of course set by the Heisenberg uncertainty principle: 
if the uncertainty $\Delta p_z$  is to be small enough for a well-defined beam to exist, 
then the width $\Delta z$ has to be at least  
${\hbar  \mathord{\left/ {\vphantom {\hbar  {\Delta p_z}}} \right. \kern-\nulldelimiterspace} {\Delta p_z}}$. 
The requirement to allow entangled wave functions is also obvious.
\footnote{Although we have seen in the literature remarks which suggest this point has been briefly lost sight of: 
	e.g. ``If we had achieved absolute separation in some region 
		(so that the tails of the two wave packets did not overlap)
		then we would have had a diagonal density matrix in that region. 
	This would have precluded attainment of a pure case, off diagonal density matrix, 
		upon recombining the beams further down-stream.''\cite{Scully:1978} 
	If achieving an absolute separation between the parts of a wave function 
		precluded subsequent interference, 
		the double-slit experiment would not work.}

We have therefore addressed two of the three problems with the collapse of the wave function. 
Except for considerations of calculational efficiency, there is no preferred basis. 
And we may give a well-defined (if virtual) location to the collapse of the wave function. 
This implies that the formalism 
-- at least for the \sg experiment 
-- is ``intuition-free'' and therefore complete in a formal sense.
(We therefore find ourselves in concurrence 
	with the position of Fuchs and Peres\cite{Fuchs:2000}: 
	``Quantum Theory Needs No `Interpretation'.'')

The next question is ``If the collapse is not necessary, why has its inclusion not done harm?''
Our answer is that it represents in most cases a very good approximation. The spatial separation of the two spin components will usually take place rapidly. 
Once this has happened, the difference between the calculation with and without collapse is minute. 
The collapse is like homeopathic medicine. 
It does no good but the difference between using it and not using it is so small it does little or no harm.

Of course homeopathic medicine, while relatively harmless, is not free. 
The collapse complicates analysis both in terms of principle -- what exactly is it? -- and in terms of practice -- at what point in the trajectory shall we invoke the collapse? 
Pending an experimental disambiguation, 
the chief arguments in favor of the collapse-free approach are that is simpler and less mysterious, 
has no adjustable parameters, 
requires the invocation of no new forces, and does not create the ``preferred basis'' problem. 
The greater conceptual simplicity of the collapse-free approach may be 
of some benefit in spintronics and quantum computation applications.

We see the collapse of the wave function as defining the boundary between the parts of the problem space which must be handled quantum mechanically and those which may be treated by classical methods. 
It is not a physical boundary -- indeed the same physical region might be treated in some ways quantum mechanically and in others classically (as we have done here, by treating the magnetic field in classical terms while treating the particles traveling through it quantum mechanically). 
The collapse is something that exists in our approximations, our software, and our minds. 
It does not, however, exist in the phenomena themselves, which are purely quantum mechanical in character.

We have only dealt with one case of the ``collapse of the wave function.''
We have not looked at the optical \sg effect\cite{Cook:1978,Cook:1987,Vaglica:1996,Chough:2001,Baym:1969} 
and we have not looked at other phenomena involving the collapse of the wave function, 
i.e. the quantum Zeno effect. 
It would be interesting to see to what extent the approach here may be generalized. 
We suggest as a general rule than whenever there is a force giving a $\Delta p$, 
the corresponding $q$ dimension should be modeled explicitly, 
as a Gaussian distribution or whatever seems appropriate.

We have also not looked at elaborations on the basic \sg idea. 
One amusing variation is to ask what might happen if the beam is sent through a ``sandwich''
of alternating layers of magnetic fields with slopes pointing in the x and z directions.
We note that the condition for the beam to split is that the impulse given to the beam 
$\mu {{\partial B} \over {\partial z }}\Delta t$
be significantly greater than the uncertainty in the momentum in the z direction of the beam
$\Delta p_z\approx {\hbar  \over {\Delta z}}={\hbar  \over \sigma }$
\cite{Baym:1969}. 
If the transition is too smooth, the beam will not be split.
But we should suddenly see a large number of small beams split off
-- each new (and sufficiently crisp) boundary should produce a new ``collapse'' 
-- if the condition  
$\mu \left| {\nabla B} \right|\gg{\hbar  \over {\sigma \Delta t}}$
is met. 
An interesting thing here is that the condition is dependent on the width of the beam; 
the wider the beam, the earlier we should see the effect.

While we have focused on the non-relativistic case the same considerations apply to quantum electrodynamics. 
The propagators are different; the principles the same.


\begin{acknowledgments}

I would like to thank Jonathan Smith for invaluable encouragement, guidance, and practical assistance.

I would like to thank Catherine Asaro, John Cramer, and Dave Kratz for helpful conversations.

I would like to thank Gaylord Ashmead, Graham Ashmead, Linda Kalb, and Ferne Welch for encouragement and support.

I would like to thank the librarians of Bryn Mawr College, Haverford College, and the University of Pennsylvania for unflagging helpfulness.

And I would like to express my appreciation of the invaluable work done by the administrators of the electronic archives at PROLA, LANL, and SLAC.

\end{acknowledgments}

\bibliography{uncollapse}

\end{document}